\documentclass[sigconf, screen]{acmart}

\usepackage{amsthm}

\usepackage{listings}
\usepackage{caption}
\usepackage{subcaption}
\usepackage[linesnumbered, ruled, vlined]{algorithm2e}
\usepackage{color}
\usepackage{xcolor}
\usepackage{multirow}
\usepackage{colortbl}
\usepackage{multicol}
\usepackage{booktabs}
\usepackage{amsfonts}
\usepackage{graphicx}
\usepackage{xcolor}
\usepackage{listings}
\usepackage{caption}
\usepackage{subcaption}
\usepackage{comment}
\usepackage{xspace}
\usepackage{url}
\usepackage{soul}
\usepackage{balance}
\usepackage{amsmath}
\usepackage{enumitem}
\usepackage{float}
\newfloat{lstfloat}{htbp}{lop}
\floatname{lstfloat}{Listing}

\newcommand\tool{RegMiner\xspace}

\newcommand\regressionNum{1035\xspace}
\newcommand\prjNum{147\xspace}
\newcommand\gtNum{50\xspace}
\newcommand\weekNum{8\xspace}

\SetCommentSty{mycommfont}

\setcopyright{acmcopyright}
\acmPrice{15.00}
\acmDOI{10.1145/3533767.3534224}
\acmYear{2022}
\copyrightyear{2022}
\acmSubmissionID{issta22main-p553-p}
\acmISBN{978-1-4503-9379-9/22/07}
\acmConference[ISSTA '22]{Proceedings of the 31st ACM SIGSOFT International Symposium on Software Testing and Analysis}{July 18--22, 2022}{Virtual, South Korea}
\acmBooktitle{Proceedings of the 31st ACM SIGSOFT International Symposium on Software Testing and Analysis (ISSTA '22), July 18--22, 2022, Virtual, South Korea}
\begin{document}

%
\title[RegMiner]{RegMiner: Towards Constructing a Large Regression Dataset from Code Evolution History}

%

\author{Xuezhi Song}
\authornotemark[2]
\affiliation{%
  \institution{Fudan University}
  \country{China}
}
\email{songxuezhi@fudan.edu.cn}

\author{Yun Lin}
\authornote{Corresponding author}
\affiliation{%
  \institution{Shanghai Jiao Tong University}
  \institution{National University of Singapore}
  \country{China/Singapore}
  }
\email{dcsliny@nus.edu.sg}

\author{Siang Hwee Ng}
\affiliation{%
  \institution{National University of Singapore}
  \country{Singapore}
  }
\email{sianghwee@u.nus.edu}

\author{Yijian Wu}
\authornote{Also affiliated with Shanghai Key Laboratory of Data Science.}
\affiliation{%
  \institution{Fudan University}
  \country{China}
  }
\email{wuyijian@fudan.edu.cn}

\author{Xin Peng}
\authornotemark[2]
\affiliation{%
  \normalsize
  \institution{Fudan University}
  \country{China}
  }
\email{pengxin@fudan.edu.cn}

\author{Jin Song Dong}
\affiliation{%
  \normalsize
  \institution{National University of Singapore}
  \country{Singapore}
  }
\email{dcsdjs@nus.edu.sg}

\author{Hong Mei}
\affiliation{%
  \normalsize
  \institution{Peking University}
  \country{China}
  }
\email{meih@pku.edu.cn}
%

%
\begin{abstract}
    Bug datasets lay significant empirical and experimental foundation for various SE/PL researches
    such as fault localization, software testing, and program repair.
    Current well-known datasets are constructed \textit{manually},
    which inevitably limits their scalability, representativeness, and the support for the emerging data-driven research.

    In this work, we propose an approach to automate the process of harvesting replicable regression bugs from the code evolution history.
    We focus on regression bugs, as they
    (1) manifest how a bug is introduced and fixed (as non-regression bugs),
    (2) support regression bug analysis, and
    (3) incorporate more specification (i.e., both the original passing version and the fixing version) than non-regression bug dataset for bug analysis.
    Technically, we address an information retrieval problem on code evolution history.
    Given a code repository, we search for regressions where a test can pass a regression-fixing commit, fail a regression-inducing commit, and pass a previous working commit.
    We address the challenges of
    (1) identifying potential regression-fixing commits from the code evolution history,
    (2) migrating the test and its code dependencies over the history, and
    (3) minimizing the compilation overhead during the regression search.
    We build our tool, \tool, which harvested \regressionNum regressions over \prjNum projects in \weekNum weeks,
    creating the largest replicable regression dataset within the shortest period, to the best of our knowledge.
    Our extensive experiments show that
    (1) \tool can construct the regression dataset with very high precision and acceptable recall, and
    (2) the constructed regression dataset is of high authenticity and diversity.
    We foresee that a continuously growing regression dataset opens many data-driven research opportunities in the SE/PL communities.
\end{abstract}

%
%


%

%
\maketitle

\section{Introduction}
Bug datasets are fundamental infrastructure to support various software engineering researches
such as software testing~\cite{boyapati2002korat, mcminn2004search, harman2009theoretical, luckow2016jd, harman2010search, lin2020recovering, lin2021graph, nguyen2020sfuzz},
fault localization~\cite{cleve2005locating, gupta2005locating, ko2008debugging, lin2017feedback, zhang2022demystifying, lin2018break}, and
bug repair~\cite{forrest2009genetic, bohme2017bug, le2016history, nguyen2013semfix, tan2015relifix}.
The SE/PL community has taken decades to construct bug datasets such as SIR~\cite{sir}, BugBench \cite{lu2005bugbench}, Codeflaws \cite{codeflaws}, and QuixBugs \cite{quixbugs},
finally gravitating to the state-of-the-art, Defects4j~\cite{just2014defects4j}, 
which takes seven years to collect over 800 bugs.

While bug datasets like Defects4j \cite{just2014defects4j} and CoREBench \cite{bohme2014corebench} have made significant contribution to the community,
there are still concerns that their sizes are relatively small (and, thus arguably less representative \cite{sobreira2018dissection}),
regarding
(1) their diversity (e.g., bug types of concurrency, API misuse, non-deterministic, etc.) and
(2) the needs of applying the emerging data-driven and AI techniques in SE tasks \cite{shin2018improving, she2019neuzz, liu2019automatic, guo2020graphcodebert, jayasundara2019treecaps}.
However, comparing to the cost of labelling an image or a sentence in the AI community,
manually collecting and labelling a bug is far more expensive.
It requires to
(1) prepare at least two versions of code project (i.e., a correct version and a buggy version),
(2) set up at least one test case passing the correct version and failing the buggy version, and
(3) isolate an environment where the bug can be well replicated.
Manually constructed bug datasets \cite{sir, just2014defects4j, bohme2014corebench, lu2005bugbench, codeflaws, quixbugs, hutchins1994experiments, liang2022gdefect4dl} naturally limit its scalability.

Recent works such as BEARS \cite{bears}, BugSwarm \cite{bugswarm}, and BugBuilder \cite{jiang2021extracting} are emerged to automate the construction of bug datasets.
BEARS \cite{bears} and BugSwarm \cite{bugswarm} collect reproducible bugs by monitoring the buggy and patched program versions from Continuous Integration (CI) system.
BugBuilder \cite{jiang2021extracting} is further designed to isolate bug-fixing changes in a bug-fixing revision.
They can generally generate a bug dataset similar to Defects4j, with a much larger size.
In such bug datasets,
the buggy code is labelled with their fixes, potentially facilitating various data-driven bug-related learning tasks.

In this work, we further propose an approach, \tool, to harvest \textit{regression bugs} from the code repository.
Regression bugs are bugs which make an existing working function fail.
Comparing to \textit{general} bug analysis (e.g., bug repair and localization) which helps to derive a passing version from a buggy version,
regression bug analysis helps to derive a passing version from a buggy version and a previously passing version.
Typical regression analysis, such as delta-debugging \cite{zelleryesterday, hodovan2016modernizing, hodovan2017coarse, kiss2018hddr, misherghi2006hdd}, reports the failure-inducing changes from a large number of changes between the buggy and the previous passing version.
The different problem setting makes Defects4j-like bug dataset hardly be useful for regression analysis.
A large regression bug dataset has the following two benefits.


\noindent\textbf{(1) Fewer Problems of Missing Specifications:}
Software bugs are essentially the inconsistencies between the implementation and its specification.
Thus, a bug-related solution such as debugging and repair can be less reliable without sufficient specification as the input.
In contrast to non-regression bug dataset where each bug is manifested with a few failing test cases,
regression bugs are additionally equipped with their past passing version
which can be used to cross-validate the violated specification with the (future) passing version,
providing a more informative semantics to
explain why the bug happens.


\noindent\textbf{(2) Scalable Benchmark for Regression Analysis:}
From the perspective of benchmark construction, a large regression dataset can lay a foundation for various regression analysis.
Existing regression analysis works (e.g., regression localization) are usually evaluated with limited number of regressions,
given that their collection is highly laborious.
Our investigation on 14 research works from the year of 1999 to 2021 \cite{yi2015synergistic, wang2019explaining, zelleryesterday, tan2015relifix, pastore2013radar, cohen2015localization, khalilian2021cgenprog, zhang2012faulttracer, yu2012towards1, yu2012towards2, brummayer2009fuzzing, yu2012practical, artho2011iterative}
shows that the mean number of evaluated real-world regressions is 16.7, the median is 12.5, and the maximum is only 40.
Moreover, different benchmarks are used in different work,
making it difficult to compare their performance.
A large-scale benchmark can not only mitigate the issue,
but support various systematic empirical studies on regressions.

In this work, we design \tool to automate the regression harvesting process with \textit{zero} human intervention.
\tool can
continuously harvest a large number of replicable regression bugs from code repositories (e.g., Github).
Technically, we address an information retrieval problem in the context of software evolution history,
i.e., retrieving runnable regressions on a code repository.
Our approach takes a set of code repositories as input, and
isolates a set of regressions with their running/replicable environment as output.
Specifically, we construct a regression by searching in code repositories for
a regression-fixing commit denoted as \textit{rfc},
a regression-inducing commit denoted as \textit{ric},
a working commit (before \textit{ric}) denoted as \textit{wc}, and
a test case denoted as $t$
so that $t$ can pass \textit{rfc} and \textit{wc}, and fail \textit{ric}.
To this end, we first design a measurement to select those bug-fixing commits with more regression potential.
Next, for each such potential regression-fixing commit, \textit{rfc},
we further search for a test case $t$ which can pass \textit{rfc},
fail the commit before (denoted as \textit{rfc}-1), and pass a commit further before \textit{rfc}-1.
Our search process addresses the technical challenges of
(1) identifying relevant code changes in \textit{rfc} to migrate through the code evolution history,
(2) adopting the library upgrades with the history, and
(3) minimizing the compilation overhead and handling incompilable revisions.

We evaluate our approach with a close-world and an open-world experiment.
In the close-world experiment, 
\tool achieves 100\% precision and 56\% recall on a benchmark consisting of \gtNum regression bugs and \gtNum non-regression bugs.
In the open-world experiment, we run \tool on \prjNum code repositories within \weekNum weeks.
\tool reports \regressionNum regressions which construct the largest Java regression dataset to the best of our knowledge.
Our ablation study also shows the effectiveness and efficiency of our technical designs.

In conclusion, we summarize our contributions as follows:
\begin{itemize}[leftmargin=*]
  \item We propose a fully automated regression mining technique,
    which allows us to continuously harvest regressions from a set of code repositories with zero human intervention.
  \item We build our \tool tool with extensive experiments evaluating its precision and recall to mine regressions.
    The results show that \tool is accurate, effective, and efficient to mine regressions from code repositories.
  \item We construct a regression dataset with \tool within \weekNum weeks. We foresee that the size of our regression dataset can keep growing with time,
      opening the a number of research opportunities on bug analysis.
\end{itemize}

The source code of \tool is available at \url{https://github.com/SongXueZhi/RegMiner}.
The mined regression dataset and its demonstration is available at \url{https://regminer.github.io/}.

\section{Problem Definition and Reformulation}\label{sec:regression-definition}


\noindent\textbf{Commit and Revision.}
A code commit can correspond to two revisions in the code repository, i.e., the revision before the commit and the revision after (or caused by) the commit.
In this work, we use the terminology \textit{commit} and its caused \textit{revision} interchangeably.
Thus, given a commit denoted by $c$, we also use $c$ to denote the revision caused by $c$.
Moreover, we use $c-1$ to denote the commit before $c$, i.e., the revision caused by the commit $c-1$.

\noindent
\textbf{Regression.}
Given a fixed regression in the code repository, we denote it as $reg = \langle rfc, ric, f \rangle$,
which indicates that the regression $reg$ consists of a regression-fixing commit \textit{rfc}, a regression-inducing commit \textit{ric}, and a feature $f$, where
(1) $f$ exists in both \textit{rfc} and \textit{ric} and
(2) $f$ works in \textit{rfc} and \textit{ric}-1 but fails in \textit{ric}.
In addition, we call the commit \textit{ric}-1 as the \textit{working commit} (WC),
and the feature $f$ as the \textit{regression feature}.

\noindent\textbf{Problem Definition.}
Formally, given a code repository $C$ as the a set of commits, we aim to construct a regression set
$REG = \{reg| reg = \langle rfc, ric, f \rangle\}$ where
$\forall reg\in REG$, $\exists f, s.t.$ $rfc\in C, ric\in C$, \textit{rfc} $\succ$ \textit{ric}, $f$ works in \textit{rfc} and \textit{ric}-1, and fails on \textit{rfc}-1 and \textit{ric}.
The operator $\succ$ indicates the chronological partial order between two commits,
and \textit{rfc} $\succ$ \textit{ric} indicates that the commit \textit{rfc} happens after the commit \textit{ric}.
The retrieval process of a fixed regression $reg$ is essentially the process of identifying \textit{rfc}, \textit{ric}, and $f$ in the code commit history.

While it is intuitive to look into a regression regarding its regression feature,
the above definition still provides limited guidance for practical implementation.
Specifically, we still need to answer the questions
(1) how do we represent a regression feature?
And more importantly, (2) how do we know that it is the same feature that works in one revision $rev_p$ in the past and fails in another revision $rev_f$ ($rev_f \succ rev_p$),
especially when $rev_f$ has gone through many changes from $rev_p$?

\noindent\textbf{Problem Reformulation (for Practical Implementation).} 
In this work, we estimate a feature with a unit test with an oracle.
The pass or failure of a unit test represents whether the feature works or fails.
Therefore, we reformulate a regression as $reg = \langle $\textit{rfc}, \textit{ric}, $t \rangle$ where $t$ is the feature-representing test.
Practically, we search for
(1) the test case $t$ and
(2) a set of regressions $REG = \{reg| reg = \langle$ \textit{rfc}, \textit{ric}, $t \rangle\}$ where
$\forall reg\in REG$, $\exists t, s.t.$ \textit{rfc}$\in C$, \textit{ric}$\in C$, \textit{rfc}$\succ$\textit{ric}, $t$ passes on \textit{rfc} and \textit{ric}-1, and fails on \textit{rfc}-1 and \textit{ric}.
Moreover, we conservatively estimate that the feature represented by $t$ in $rev_p$ (i.e., passing revision)
\textbf{\textit{is compatible with}} that in $rev_f$ (i.e., failing revision) in practice if:
\begin{enumerate}[leftmargin=*]
  \item The methods tested by $t$ in \textit{rfc} and \textit{ric} are similar\footnote{In practice, we consider two methods are similar if (1) their method names are the same and (2) the similarity between code bodies is above a threshold $th_{body}$ (e.g., 0.95).}.
  \item The assertion failures of $t$ in \textit{rfc}-1 and \textit{ric} share the same root cause\footnote{We regard they share the same root cause if they share the same error message and error-occurring location in $t$.}.
\end{enumerate}


We adopt the conservative estimation because our regression mining approach generally favours precision over recall.
Given the test $t$ in a regression $\langle rfc, ric, t \rangle$, we also say that
(1) $t$ is the \textit{regression test} and
(2) $t$ can \textit{compatibly} pass \textit{rfc} and \textit{ric}-1, and fail \textit{rfc}-1 and \textit{ric}.

\section{Overview}

We design our approach to mine and construct a regression dataset as Algorithm~\ref{alg:overview},
which takes as input a set of code repositories and a similarity threshold to select potential regression-fixing commits.
When searching for the regressions,
our approach first collects the bug-fixing commits including test case addition from the code repositories (line 3-6).
We further confirm a commit $c$ as bug-fixing if the added test case can pass $c$ and fail $c-1$ (line 5-6),
which serves as a prerequisite to search the regression bug.
Then we estimate $c$'s potential as a regression-fixing commit (line 7).
With the quantified regression potential,
we can rank the bug-fixing commits and remove those with less potential (line 10).
For each potential regression-fixing commit,
we search for its regression-inducing commit \textit{ric} which satisfies that the test $test$ can \textit{compatibly} fail in \textit{ric} and pass in \textit{ric}-1 (line 12).
Then, we record a regression $reg$ by \textit{rfc}, \textit{ric}, and the relevant test case $test$.
The design of Algorithm~\ref{alg:overview} needs to overcome the following three challenges.
\begin{algorithm}[t]
    \SetNoFillComment
    \small
    \caption{Regression Dataset Construction}\label{alg:overview}
    \SetKwInOut{Input}{Input}
    \SetKwInOut{Output}{Output}
    \Indm

    \Input{A code repository set, $repositories$; a threshold of regression potential, $th_{rp}$}

    \Output{A regression dataset, $regressions$}
    \Indp
    \BlankLine

    \tcp{initialize the regression set}
    $regressions = \emptyset$

    \For{$repo \in repositories$}{
        \tcp{initialize the regression set}
        $commits =$ \texttt{search\_commits\_with\_test\_addition}($repo$)

        \For{$c \in commits$}{
            \tcp{fixing commit confirmation}
            $is\_fix =$\texttt{confirm\_fix}($c.test, c, c-1$)

            \If{$is\_fix$}{
                \tcp{RFC prediction}
                \texttt{estimate\_regression\_potential}($c$)
            }
            \Else{
                $commits = commits \setminus \{c\}$
            }
        }

        $commits' =$\texttt{rank\_and\_filter}($commits, th_{rp}$)

        \For{$c \in commits'$}{
            \tcp{search regression with test migration}
            $ric, test =$\texttt{search\_regression}($c.tests, repo$)

            \If{\textit{ric} != null}{
                $rfc = c$

                $reg = (rfc, ric, test)$

                $regressions = regressions \cup \{reg\}$
            }

        }
    }

    \Return $regressions$
\end{algorithm}

\noindent\textbf{Challenge 1: Futile Search on Non-regression Bug-fixing Commits.}
Searching for a regression across the commit history is time-consuming, which includes the overhead of
project compilation,
test case migration, and
test case execution.
Starting with a non-regression fixing commit causes the whole search futile.
In this work, we design a novel measurement to quantify the potential of a bug-fixing commit to be a regression-fixing commit.



\noindent\textbf{Challenge 2: Test Dependency Migration.}
With a test case and the potential regression-fixing commit, 
it is non-trivial to verify its regression-inducing commit.
In practice, the regression can be induced years ago.
The project can undergo radical changes and depend on different versions of libraries.
Without appropriate adaption, we can miss a large number of real regressions.

\noindent\textbf{Challenge 3: Large Overhead to Validate Regression.}
Starting from a potential regression-fixing commit, there could be thousands of commits to check out, recompile, and run.
Sequentially checking out revisions incurs huge runtime overhead,
while bi-sect approach can suspend on some incompilable commits.


\section{Search Methodology}\label{sec:approach}
In this section, 
we introduce how we address the three aforementioned challenges.

\subsection{Estimating Regression Potential}
Let us call the set of code elements implementing the regression feature as \textit{feature code}.
Assume that we know that
(1) a feature $f$ is fixed in a revision \textit{bfc}, and
(2) the precise set of code elements (e.g., methods) $set_f$ of feature code,
we can estimate the probability of \textit{bfc} to be a regression-fixing commit as
\begin{equation}\label{eq:probability}
  P(rfc | bfc, set_f) = 1 - (1-p)^N
\end{equation}

In \autoref{eq:probability},
$N$ is the number of changes applied on $set_f$ in the commits before \textit{bfc},
$p$ is the probability that a change introduces a bug on the feature $f$, which we call as \textit{regression-inducing probability}.
The more changes applied on $set_f$ in the history, the more likely the \textit{bfc} is a regression-fixing revision.

However, it is non-trivial to precisely infer the program elements $set_f$ purely based on a test case. 
Including irrelevant elements in $set_f$ can make non-regression fixing revisions have a larger probability.
In contrast, missing important elements in $set_f$ can make real regression-fixing revision have a lower probability.
Either case can make \autoref{eq:probability} misguide the prediction of regression-fixing commit.

\subsubsection{Feature Code Identification}
We measure the relevance of a method $m$ to a test case $t$ by its uniqueness and similarity to $t$:
\begin{equation}\label{eq:method-relevance}
  rel(m) = min(1.0, test\_uniq(m, t) \times (1 + sim(m, t)))
\end{equation}

In \autoref{eq:method-relevance}, the relevance score $rel(m)$ depends on
(1) its test uniqueness ($test\_uniq(m, t)$), i.e., how unique $m$ is to $t$ and
(2) the textual similarity between $m$ and $t$ ($sim(m, t)$).
We design $test\_uniq(m)$ and $sim(m, t)$ with the range [0, 1].
The $min(.)$ function in \autoref{eq:method-relevance} ensures that the relevance score has an upper bound of 1.0.

\paragraph{Test Uniqueness}
To evaluate the test uniqueness of a method $m$ for $test$, we design a variant TF-IDF measurement.
Assume that there are $N$ test cases in a project, and each test case has many methods in its call hierarchy.
Thus, we can construct a test-method matrix $R_{m, t}$ where each column represents a test case and each row represents a method called by at least one test case.
Specifically, we calculate an IDF-like measurement for each $r(m, t)$ in $R_{m, t}$ as
\begin{equation}\label{eq:idf}
  test\_uniq(m, t) = r(m, t) = \left\{
                 \begin{array}{ll}
                   log_{N}{\frac{N}{freq(m)}}  & \mbox{if $t$ covers $m$}\\
                   0 & \mbox{otherwise}\\
                 \end{array}
               \right.
\end{equation}

In \autoref{eq:idf}, $N$ represents the number of test cases in the project,
$freq(m)$ represents the number of test cases covering method $m$, which ranges from 1 to $N$.
Thus, $log_{N}{\frac{N}{freq(m)}}$ is scaled to the range [0, 1].

\paragraph{Textual Similarity}
Moreover, we further introduce a similarity function between $t$ and $m$.
Specifically, we tokenize the name of test method into a bag of words with the word ``test'' removed, e.g.,
\text{testCalendarTimeZoneRespected} is converted into a bag as $B=$\{``calendar'', ``time'', ``zone'', ``respected''\}.
Let the number of tokens in the name/body of $m$ which can match any of words in $B$, be $k$,
then the $sim(test, m) = \frac{k}{|B|}$, which ranges between 0 and 1.




\subsubsection{Code Element Re-identification}\label{sec:code-reidentification}
To collect historical modifications on a method,
we need to \textit{re-identify} them in the past commits.
Given a method $m$ in a revision $r$, if we can find a method $m'$ in another revision $r'$ with exact the same signature as $m$ ($r' \succ r$),
we consider $m'$ as a match of $m$ in $r'$.
If we cannot locate such an exact match,
we track their identity by defining their similarity:
\begin{equation}\label{eq:method-similarity}
  sim(m, m') = \alpha \cdot sim\_signature(m, m') + \beta \cdot sim\_body(m, m')
\end{equation}

The metrics consist of the signature similarity and the code body similarity.
We consider $m$ and $m'$ as a match if there similarity is above a threshold $th_m$.
The similarity metrics are generally defined regarding method name similarity (the ratio of longest common subsequence over length of two method names), return types (same type or not), parameter types (Jaccard coefficient of two parameter type set), and method bodies (the ratio of longest common subsequence over length of two method bodies).
Readers can refer to more details in our anonymous website \cite{regminer}.

\subsubsection{Regression Potential Metric}
Consequently, given a set of methods $M$, each $m\in M$ has a relevance score denoted by $rel(m)$ and a historical change number $m.changes$.
Thus, we quantify the final probability of a regression-fixing commit as
\begin{equation}\label{eq:practical-measurement}
  P(rfc | bfc, set_f) = 1 - \prod_{m\in M} (1-p \times rel(m))^{m.changes}
\end{equation}
In \autoref{eq:practical-measurement},
$p$ is the base empirical regression-inducing probability, e.g., 0.01, shared by all the methods.
\subsection{Test Migration}\label{sec:test-dependency-migration}

%
%
%
%
%

Given a test $t$ in a bug-fixing commit \textit{bfc},
we need to migrate $t$ along with its code dependency in a revision $c_{inv}$ under investigation ($bfc \succ c_{inv}$).
Typically, we need to overcome two challenges to avoid compilation errors:
\begin{itemize}[leftmargin=*]
  \item \textbf{Identifying migrating dependencies.}
    When we migrate $t$ to the revision $c_{inv}$,
    the code dependencies of $t$ should be available in $c_{inv}$.
  \item \textbf{Reconciliating migrated code.}
    The migrated test and its dependencies can be adapted to the old libraries.
\end{itemize}

\subsubsection{Identifying Migrating Dependencies}
We consider a \textit{code element} as either a class, interface, field, or method.
We aim to find a minimum set of code elements in the bug-fixing revision \textit{bfc} so that
(1) the revision under investigation $c_{inv}$ migrated with the test $t$ can still compile and run, and
(2) the migration does not change the program behaviors of test case $t$ under $c_{inv}$.
We achieve the former by parsing the minimum code elements depended by the regression test, and
achieve the latter by identifying and refraining the migration of code elements with the bug-fixing changes.
\autoref{fig:migrating-workflow} shows the workflow of identifying the code elements to be migrated,
which consists of the following steps:

\begin{figure}[h]
  \centering
  \includegraphics[scale=0.42]{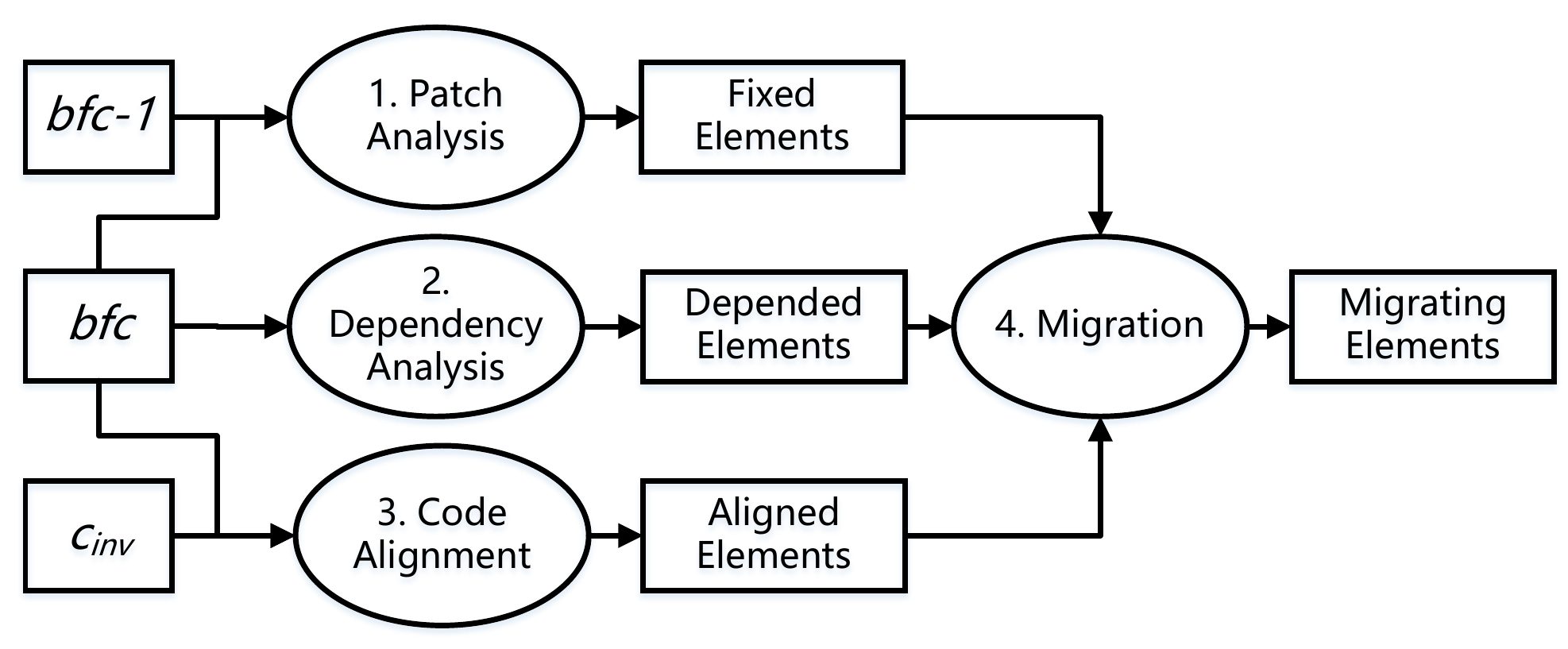}
  \caption{Identifying Dependency to be Migrated}\label{fig:migrating-workflow}
\end{figure}

\begin{figure*}[t]
  \centering
  \includegraphics[scale=0.43]{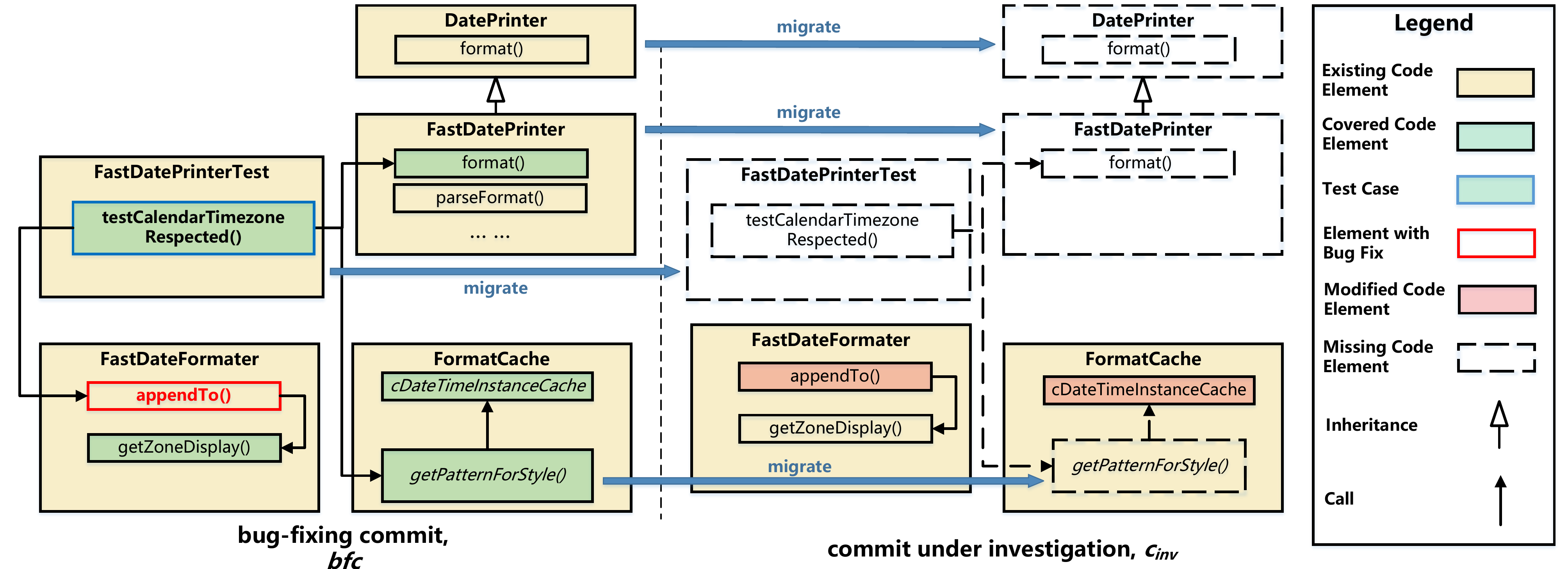}
  \caption{Migrating test dependencies to a revision.
  }\label{fig:migrating-example}
\end{figure*}

\noindent\textbf{Step 1. Patch Analysis.}
We compare the revisions \textit{bfc} (where $t$ passes) and \textit{bfc}-1 (where $t$ fails) to locate the code elements in \textit{bfc} where the fix happens.
We denote such set of elements as $E_{fix}$.
Readers can refer to \cite{jiang2021extracting} for more details.

\noindent\textbf{Step 2. Dependency Analysis. }
Still in \textit{bfc}, we calculate the call graph rooted at the test case $t$ (denoted as $E_{cg}$) and the set of covered elements by running $t$ (denoted as $E_{ce}$).
We keep the minimum dependencies in $E_{cg}$ and $E_{ce}$.

\noindent\textbf{Step 3. Code Alignment. }
Given \textit{bfc} and the revision under investigation $c_{inv}$, we detect how the code are aligned between two revisions.
Thus, we can avoid migrating a code element matchable in $c_{inv}$.
We use Xing and Stroulia's UMLDiff algorithm \cite{xing2005umldiff} to align the cross-revision code elements.
As a result, we use $E_{id}$ to denote the set of the identical elements between two revisions, and
$E_{mod}$ to denote the set of the modified elements (i.e., the elements matched in \textit{bfc} with $c_{inv}$ but with difference).

\noindent\textbf{Step 4. Migration. }
Finally, the set of migrating elements is calculated as the missing dependencies $E_{miss} = (E_{cg} \cup E_{ce}) \setminus (E_{fix} \cup E_{id})$,
and the changed dependencies $E_{change} = (E_{cg} \cup E_{ce}) \cap (E_{mod} \setminus E_{fix})$.
We will reconcile $E_{mig} = E_{miss} \cup E_{change}$ on $c_{inv}$.
We will elaborate the reconciliation with more details in Section~\ref{sec:reconciliation}.

\paragraph{Example}
\autoref{fig:migrating-example} shows simplified class dependency graphs of a regression-fixing commit \textit{bfc} and
a commit under investigation $c_{inv}$ in Common Lang project.
In \autoref{fig:migrating-example},
we use yellow container boxes to represent classes (or interfaces), and
containee boxes to represent class members such as method and field.
The containment relations represent that a class/interface declares a method or a field.
A method is a containee box whose name has ``()'' as suffix, while fields are containee boxes without the suffix.
The black arrows between the boxes are the call relation between methods and fields.
We represent in green the elements covered by the test case in the bug-fixing commit.

In addition, the position of each box in \textit{bfc} and $c_{inv}$ represents the alignment relation between two revisions.
For example, the class \texttt{FastDataFormater} in \textit{bfc} is aligned with the class with the same name in $c_{inv}$.
The missing depended elements by $t$ in $c_{inv}$ is represented by dashed boxes,
the modified elements in $c_{inv}$ is represented by red boxes, and
the elements with code fix in \textit{bfc} is represented by boxes with red border.

The workflow to identify the migration of code elements in \textit{bfc} as follows.

\noindent\textbf{1. Patch Analysis.}
In \textit{bfc}, we first identify the fixes by comparing \textit{bfc} and \textit{bfc}-1
to have $E_{fix} = \{$\texttt{Fast\-Date\-Formater.\-appendTo()}$\}$.

\noindent\textbf{2. Dependency Analysis.}
Next, the static and dynamic dependency analysis from the test case \texttt{test\-Calendar\-Timezone\-Res\-pect\-ed()} produce the set $E_{cg} \cup E_{ce} = \{$
\texttt{test\-Calendar\-Timezone\-Res\-pect\-ed()},
\texttt{Fast\-Date\-Printer.\-format()},
\texttt{Date\-Printer.\-format()},
\texttt{Fast\-Date\-Formater.\-append\-To()},
...
$\}$ (i.e., all the green boxes and the red-bordered box in \textit{bfc}).

\noindent\textbf{3. Code Alignment.}
Then, we align two revision \textit{bfc} and $c_{inv}$ to know what dependencies of the test case are missing or changed.
For example, we can know the missing dependencies such as \texttt{Fast\-Date\-Printer.\-format()} and
\texttt{Date\-Printer.\-format()}.
Also, we know that the field \texttt{cDateTimeInstanceCache} (in red) is modified and has a potential to reconcile in $c_{inv}$.

\noindent\textbf{3. Migration.}
Finally, we calculate the missing dependencies as $E_{miss} = \{$
\texttt{test\-Calendar\-Timezone\-Respect\-ed()},
\texttt{Fast\-Date\-Prin\-ter.\-for\-mat()},
\texttt{Date\-Printer.\-format()},
\texttt{Format\-Cache.\-get\-Pa\-tt\-ern\-For\-Style()}
$\}$, and the changed dependencies as $E_{change} =\{$
\texttt{Format\-Cache.cDate\-Time\-Instance\-Cache}
$\}$.
Note that, the code elements with fix (i.e., \texttt{FastDateFormater.appendTo()}) is excluded.

As for $E_{mig} = E_{miss} \cup E_{change}$,
we define transformation rules to adapt them in $c_{inv}$.

\subsubsection{Reconciliating Migrated Code}\label{sec:reconciliation}
We design our reconciliation heuristics with the following principles:
\begin{itemize}[leftmargin=*]
  \item \textbf{Missing Dependencies:}
    The code elements in $E_{miss}$ are copied directly to $c_{inv}$ without modifications.
    The copied $E_{miss}$ will be adapted only if their migration incurs compilation errors on $c_{inv}$.
  \item \textbf{Changed Dependencies:}
    The elements in $E_{change}$ are not migrated unless their migration can
    minimize the compilation errors caused by the migration.
\end{itemize}

\begin{algorithm}[t]
    \SetNoFillComment
    \small
    \caption{Migration Reconciliation}\label{alg:reconciliation}
    \SetKwInOut{Input}{Input}
    \SetKwInOut{Output}{Output}
    \Indm

    \Input{missing dependencies, $E_{miss}$;
    changed dependencies, $E_{change}$;
    revision under investigation, $c_{inv}$;
    regression test, $test$;
    }

    \Output{A migrated revision, $c'_{inv}$}
    \Indp
    \BlankLine

    $c'_{inv} = migrate(E_{miss}, c_{inv})$

    \If{$compatible(c'_{inv}, test)$}{
        \Return $c'_{inv}$
    }
    \Else{
        $c'_{inv0} = transform(c'_{inv})$

        \If{$compatible(c'_{inv0}, test)$}{
            \Return $c'_{inv0}$
        }
        \Else{
            $c'_{inv1} = migrate(E_{change}, c'_{inv0}, test)$

            \If{$compatible(c'_{inv1}, test)$}{
                \Return $c'_{inv1}$
            }
            \Else{
                $c'_{inv2} = transform(c'_{inv1}, test)$

                \If{$compatible(c'_{inv2}, test)$}{
                    \Return $c'_{inv2}$
                }
            }
        }
    }

    \Return null
\end{algorithm}

Algorithm~\ref{alg:reconciliation} shows our design to reconcile migration,
which takes as input the missing dependencies $E_{miss}$, the changed dependencies $E_{change}$, the revision under investigation $c_{inv}$, and the regression test $test$;
and derives a migrated revision to compatibly run the regression test.
Generally, Algorithm~\ref{alg:reconciliation} is designed as a decision tree where each decision node guides the follow-up operations.
Specifically, the decision is on
whether the migrated revision is compilable.
In Algorithm~\ref{alg:reconciliation}, the conditions are denoted in line 2, 6, 10, and 14.
If the revision under transformation conforms to the condition, we return it as the revision;
otherwise, we proceed to the next branch until either we can find a revision satisfying the condition or the migration is aborted (line 16).

In Algorithm~\ref{alg:reconciliation}, we first only migrate missing dependencies $E_{miss}$ (line 1).
If not successful, we proceed to transform the migrated revision with our pre-defined AST rewriting rules.

The rewriting rules transform the code regarding the different library versions in $c_{inv}$ and \textit{bfc} (line 5).
In the transformation rules,
we set a list of triggers such as syntax change and library version change.
Those rules are generally defined regarding the API changes between the versions.
\autoref{eq:rewriting-rule} shows one of our defined rules, in the form of the operational semantics.
The precondition (the numerator) indicates that the JDK versions are different,
and an existing to-be-migrated method ($m\in M_{mig}$) defined in an interface ($m\in M_{inf}$) has an annotation of ``@Override'' ($``@Override"\in m.ANN$).
The postcondition (the denominator) indicates that we remove the ``@Override'' for that method.
For example in \autoref{fig:migrating-example},
when detecting that the supported JDK version in $c_{inv}$ is 1.5 while that in \textit{bfc} is 1.7,
we will remove the ``Override'' annotation on \texttt{format()} method in \texttt{FastDatePrinter}\footnote{In JDK 1.5, the ``Override'' annotation is not allowed to describe a method implementing the interface}.
Readers may refer to \cite{regminer} for more details.

\begin{equation}\label{eq:rewriting-rule}
  \frac{
  \begin{split}
     JDK\_v(c_{inv}) = 1.5, & JDK\_v(bfc)>1.5, \\
       & \exists m\in M_{inf} \wedge m\in M_{mig}, ``@Override"\in m.ANN
  \end{split}
  }
  {m.ANN \rightarrow m.ANN\setminus \{``@Override"\}}
\end{equation}

We proceed to migrate the changed dependencies if the transformation is not successful (line 8).
Different from migrating missing dependencies, we migrate the changed dependencies as an optimization problem.
We aim to migrate and transform the minimum changed dependencies to derive a compatible revision (line 9 and 13).
Specifically, we repetitively overwrite the code elements in $c_{inv}$ with elements in $E_{change}$ to search for a minimum core to satisfy the conditions.
In the example of \autoref{fig:migrating-example},
we will overwrite the field \texttt{cDateTimeInstanceCache} in $c_{inv}$ with that in \textit{bfc} for eradicating the compilation errors.
Note that, the static method \texttt{get\-Pattern\-For\-Style()} invoking that field is migrated to $c_{inv}$, which leads to the change of the field's modifier to \textit{static}.

\subsection{Validation Effort Minimization}\label{sec:validation-effort-minimization}
Given a bug-fixing commit, we validate it as a regression-fixing commit by searching for a regression-inducing commit $c$ in the history
where the test case fails $c$ but passes $c-1$.
A naive search algorithm can be a binary search algorithm as \textit{git bisect} implementation \cite{git-bisect}.
The binary search algorithm assumes that each revision provides us with a feedback (such as test case pass or failure),
which guides us to either approach
the revision $c$ where $c$ can fail and $c-1$ can pass (i.e., regression bug), or
the initial revision where $c$ just fails (i.e., non-regression bug).

However, the revisions cannot always be compilable in practice.
Moreover, the test code and its dependencies migrated by \tool can also suffer from incompilability.
In such a case,
the potential compilation errors provide less guidance
and the test case cannot be resolved to provide feedback.
Therefore, when visiting a revision with no feedback during the search,
we design an approach to either
(1) search for the closest revision with guiding feedback to get rid of the ``no-feedback region'' in the history, or
(2) quickly abandon the bug-fixing commit to proceed with the next one.

\begin{figure}[t]
  \centering
  \includegraphics[scale=0.5]{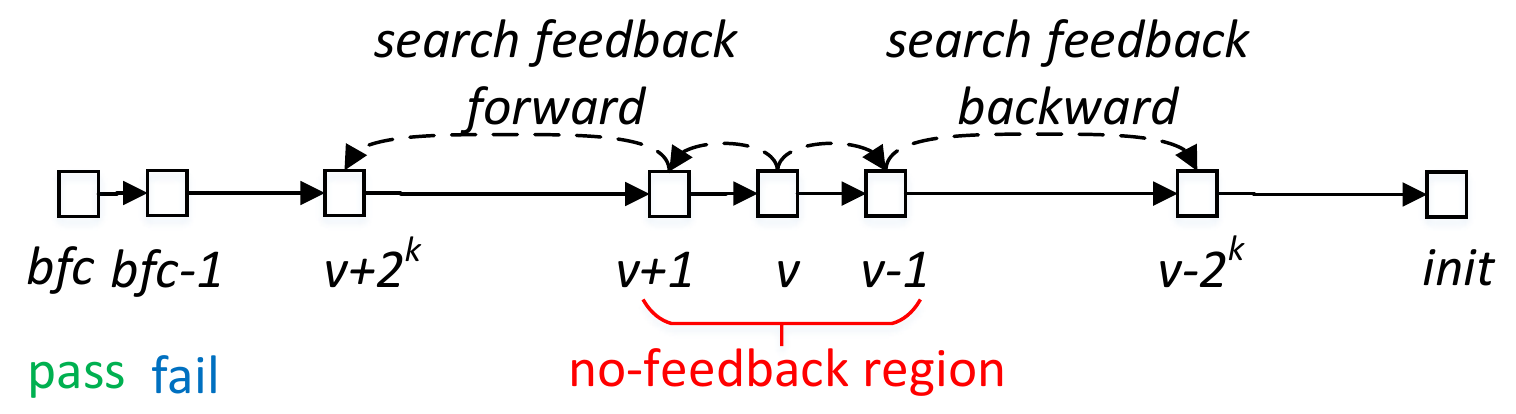}
  \caption{Search feedback revisions forward and backward}\label{fig:regression-search}
\end{figure}

\paragraph{Skipping No-feedback Region}
\autoref{fig:regression-search} shows an example where we visit a no-feedback revision when searching over the code history.
Our approach estimates the no-feedback region with an exponential search algorithm as Algorithm~\ref{alg:no-feedback-region}.
Taking a no-feedback revision $v$ and a boundary revision as input $b$ (indicating the searched boundary region cannot be beyond ($v$, $b$) or ($b$, $v$)),
Algorithm~\ref{alg:no-feedback-region} returns a revision $b_r$ between $v$ and $b$ so that $b_r$ can be either
(1) a feedback revision closest to the no-feedback revision $v$ or
(2) the boundary revision $b$ if no such feedback revision can be found.
Specifically, each time we move on the history, we double the step size to get rid of the no-feedback region soonest possible (line 21).
When we reach a feedback revision from a non-feedback revision (line 8-11) or vice versa (line 19-20),
we change the search direction and reset the step size as 1 to fine-tune the region.
The boundary revision closest to the initial no-feedback revision $v$ will be preserved (line 9 and 15).
Finally, the optimal $b_r$ will be returned if the search on the history is out of pre-defined scope (line 13 and 17).

\begin{algorithm}[t]
    \SetNoFillComment
    \small
    \caption{Closest Feedback Revision Search}\label{alg:no-feedback-region}
    \SetKwInOut{Input}{Input}
    \SetKwInOut{Output}{Output}
    \Indm

    \Input{A no-feedback revision, $v$;
    a code repository, $repo$;
    a test case, $test$;
    boundary revision $b$}

    \Output{A boundary revision with feedback, $b_r$}
    \Indp
    \BlankLine
    \tcp{the search direction is initialized as from $v$ towards $b$}
    $direction = v \succ b$

    $cursor = prev = v$

    \tcp{the revision to be returned}
    $b_r = b$

    \tcp{the search range is within $b_{past}$ and $b_{future}$}
    $b_{future} = max(v, b), b_{past} = min(v, b)$

    $step = 1$

    \While{true}{

        $prev = cursor$

        \tcp{move towards the direction with a step size}
        $cursor = $ move$(cursor, step, direction, test)$

        \tcp{when we find a revision $b_r$ which can first feedback}
        \If{$cursor$ has feedback and $prev$ has not feedback}{
            $b_r$ = update\_best($b_r$, $cursor$)

            $direction = \neg direction$

            $step = 1$
        }
        \ElseIf{$cursor$ has feedback and $prev$ has feedback}{
            \If{$cursor$ is out of boundary $(b_{past}, b_{future})$}{
                \Return $b_r$
            }
            $b_r$ = update\_best($b_r$, $cursor$)
        }
        \ElseIf{$cursor$ has not feedback and $prev$ has not feedback}{
            \If{$cursor$ is out of boundary $(b_{past}, b_{future})$}{
                \Return $b_r$
            }
        }
        \ElseIf{$cursor$ has not feedback and $prev$ has feedback}{
            $cursor$ = $prev$

            $step = 1$
        }

        $step = step \times 2$
    }

\end{algorithm}

\paragraph{Overall Algorithm}

Taking as input a revision $head$ and a revision $tail$ where $tail \succ head$, a test case $test$, and the code repository $repo$,
Algorithm~\ref{alg:regression-introduction-search} returns a regression-inducing revision, if exists.
Algorithm~\ref{alg:regression-introduction-search} is designed based on binary search (line 1-7, 20).
However, if we visit a no-feedback revision during the binary research,
we will search for the no-feedback region supported by Algorithm~\ref{alg:no-feedback-region} (line 9 and 10).
Based on the reported region, we either
(1) skip this bug-fixing commit (line 11-12, i.e., cannot find a feedback revision between $head$ and $tail$), or
(2) reset the binary search region to continue the search.

\begin{algorithm}[t]
    \SetNoFillComment
    \small
    \caption{Regression Introduction Search}\label{alg:regression-introduction-search}
    \SetKwInOut{Input}{Input}
    \SetKwInOut{Output}{Output}
    \Indm

    \Input{A revision, $head$; A revision, $tail$; a code repository, $repo$; a test case, $test$}

    \Output{A regression-inducing revision, $ric$}
    \Indp
    \BlankLine

    $v$ = $repo$.get\_middle$(head, tail, test)$



    \While{$tail \succ head$}{
        \If{$v$ has feedback}{
            \If{$v$ can pass}{
                $head = v$
            }
            \Else{
                $tail = v$
            }
        }
        \Else{
            $b_1$ = search\_feedback\_revision($v$, $repo$, $test$, $head$)

            $b_2$ = search\_feedback\_revision($v$, $repo$, $test$, $tail$)


            \If{$head == b_1 || tail == b_2$}{
                \Return null;
            }

            \If{$b_2$ can pass}{
                $head = b_2$;
            }
            \ElseIf{$b_1$ can pass and $b_2$ can fail}{
                $head = b_1$;

                $tail = b_2$;
            }
            \ElseIf{$b_1$ can fail and $b_2$ can fail}{
                $tail = b_1$;
            }

        }

        $v$ = $repo$.get\_middle$(head, tail, test)$

    }

    \Return $head+1$
\end{algorithm}

\section{Experiment}\label{sec:evaluation}
We build \tool to mine Java regressions, supporting maven and gradle projects.
We evaluate \tool with the following research questions,
more details of our experiment are available at our anonymous website \cite{regminer}.
\begin{itemize}[leftmargin=*]
  \item \textbf{RQ1 (Close-world Experiment):} Can \tool accurately and completely mine regressions from Git repositories?
  \item \textbf{RQ2 (Open-world Experiment):} Can \tool continuously mine authentic regressions from real-world Git repositories? How diverse is the regression dataset?
  \item \textbf{RQ3 (Ablation Study):} How does each technical component in \tool
    contribute to the mining effectiveness and efficiency?
\end{itemize}

\subsection{Close-world Experiment}\label{sec:rq1}
\subsubsection{Experiment Setup}
We manually collect \gtNum regression-fixing commits and \gtNum non-regression fixing commits
from Github repositories for the measurement of precision and recall.
To avoid bias, we construct the regression benchmark without any help of \tool.
Specifically, we search for closed Github issues each of which uniquely mentions a commit as its solution.
Then, we filter those issues based on its labels and description.
We prioritize the issues with labels as ``regression'' or ``bug'', or with description with the keyword of ``regression''.
Then, we confirm the real regressions by manually checking the evolution history for regression-inducing commits, and migrating the test cases.
As a result, we confirmed \gtNum regressions from 23 Java projects (see \cite{regminer} for more details).
By similar means, we identified \gtNum non-regression bugs.

We apply \tool on all the regression and non-regression fixing commits to measure the precision and recall.
Let the number of true regressions be $N$ ($N=50$), the number of reported regressions be $M$, and the number of reported true regressions be $K$,
the precision is $\frac{K}{M}$, the recall is $\frac{K}{N}$.
We choose the empirical regression-inducing probability $p$ as 0.05.

\subsubsection{Results}
\tool achieves 100\% precision and 56\% recall in the experiment.
Since the number of selected non-regression fixing commit is small in this experiment,
we leave the discussion on the potential false positives to Section~\ref{sec:rq2}.
Next, we discuss the reasons of false negatives.
Our investigation shows that the recall mainly suffers from three folds:
(1) our engineering implementation has not supported some sophisticated project configurations (16 out of 22),
(2) insufficient AST adaption (see Section~\ref{sec:test-dependency-migration}) for test case migration (5 out of 22), and
(3) our validation effort minimization can sometimes miss the regression-inducing commit (1 out of 22),

\begin{figure}[t]
  \centering
  \includegraphics[scale=0.5]{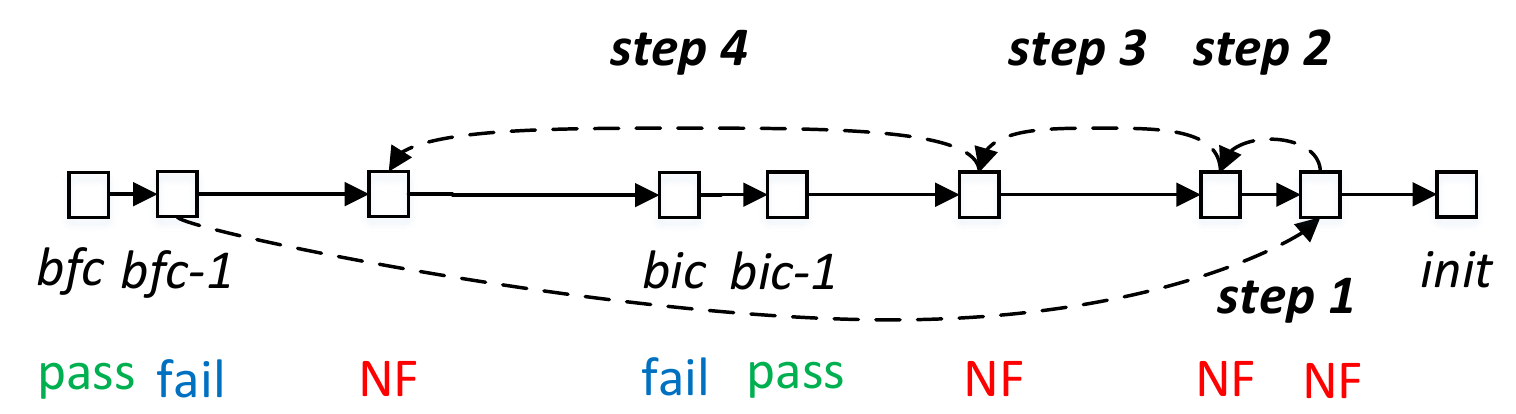}
  \caption{Validation effort minimization may cause the false negatives, NF represents the revision without feedback.}\label{fig:search-failure-1}
\end{figure}

\paragraph{Limited Engineering Support and Test Dependency Migration}
Our investigation on the false negatives reveals that
the realization of our methodologies requires more sophisticated implementation beyond our current engineering support.
\tool still has its limitation under the scenarios such as
the project-build configuration beyond maven and gradle,
specific maven plugin (e.g., maven license plugin can stop the project building if our migration does not include licence information in the source code), and
the dependency on GCC to compile the project.
Moreover, \tool also has its limitation to handle complicated Java syntax like lamba expressions or \texttt{enumerate} classes.
We will strengthen our engineering support to generalize \tool in more practical scenarios.

\paragraph{Validation Effort Minimization}
We observe that, the efficiency pursued by our VEM (Validation Effort Minimization) technique, sometimes can scarify the accuracy, leading to the false negatives.
\autoref{fig:search-failure-1} shows an example.
In \autoref{fig:search-failure-1}, we show the visiting order of \tool on one ground truth regression (\texttt{univocity-parsers} project with fixing commit \texttt{e48cdc}).
In this case, \tool visits non-feedback revisions in all the first four steps.
Based on Algorithm~\ref{alg:regression-introduction-search}, \tool tries to search for a feedback revision by
exponentially increasing the search step size (line 23 in Algorithm~\ref{alg:no-feedback-region}).
A mitigation is to change the exponential increase to a less radical increase approach, e.g., increase the moving step by a fix size.
Nevertheless, it will incur large overhead as a tradeoff.
We leave the choice to the practitioners in their own applications.

\subsection{Open-world Experiment}
\subsubsection{Experiment Setup}
In this experiment, we collect Git repositories from Github and run \tool to mine regressions for \weekNum weeks.
We evaluate \tool regarding the authenticity and the diversity of
the mined regressions in both quantitative and qualitative manner.

\paragraph{Quantitative Analysis}
We quantitatively estimate how likely the tested feature in the regression-fixing revision (RFC) exists in the regression-inducing revision (RIC),
based on the rationale that the feature in RFC is likely to be preserved in RIC
if the test covers similar set of code elements in both revisions.
Specifically, for each mined regression $reg = \langle rfc, ric, t \rangle$,
we run $t$ against \textit{rfc} and \textit{ric} to have their covered method set $E_{rfc}$ and $E_{ric}$.
We consider $m_f \in E_{rfc}$ can be aligned with $m_i \in E_{ric}$ if UMLDiff algorithm ~\cite{xing2005umldiff} can match them.
Therefore, we measure the feature coverage similarity in \textit{rfc} and \textit{ric} by:
$$
sim_f(reg) = \frac{|E_{rfc} \cap E_{ric}|}{min(|E_{rfc}|, |E_{ric}|)}
$$


\paragraph{Qualitative Analysis}
We further sample 30 regressions with high coverage score (>0.9), 30 with medium score (0.4-0.7), and 30 with low score (<0.1).
To facilitate the manual investigation, we build a tool to visualize and compare revision difference, run test cases against each revision, and revert changes between revisions.
A screenshot of the tool is available at our website \cite{regminer}.
We recruit two graduate students (both with over 5 years of Java programming experience) to manually verify each regression based on the following template questions:
\begin{itemize}[leftmargin=*]
  \item What is the feature in RFC aims to fix?
  \item What is the reason of the regression bug?
  \item Why can the feature work well before RIC?
\end{itemize}
The two students were asked to write down their answer independently.
Then, they are further asked to discuss with each other to reconcile their answers.
Given a regression, if they agree that the feature tested in RFC is different from the feature tested in RIC,
we conclude the regression is a false positive.

\paragraph{Diversity Analysis}
We measure the diversity of the regression dataset by investigating the diversity of
(1) topics of used libraries of the regression and
(2) the exception types of each bug.
Assume the regressions are distributed in $K$ library topics\footnote{We follow the topic categories defined in maven central repository (i.e., \url{https://mvnrepository.com/repos/central}), e.g., core utilities, collection, JSON libraries, etc.}, and reported with $L$ exception types (e.g., \texttt{NullPointerException}),
we compare the diversity between our collected datasets and Defects4j on
(1) the number of library topics and reported exception types, and
(2) the information entropy of bug distribution on the topics and exception types.

\subsubsection{Results}\label{sec:rq2}
In this experiment, we explore 30,165 commits in which 6,253 commits are bug-fixing commits.
As a result, \tool constructs a regression dataset consisting of \regressionNum regressions over the \prjNum projects within \weekNum weeks.
Among the \regressionNum mined regressions, over 72\% commits incur runtime exceptions when the test is migrated to a past commit.
Nevertheless, \tool still manages to report them with the validation effort minimization technique (see Section~\ref{sec:validation-effort-minimization}).

\begin{figure}
  \centering
  \includegraphics[scale=0.25]{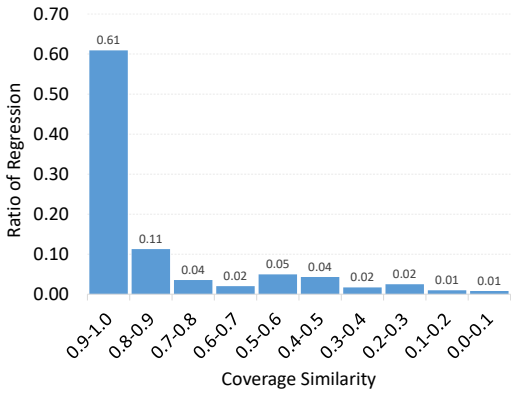}
  \caption{Feature Coverage Similarity Distribution}\label{fig:coverage-distribution}
\end{figure}

\paragraph{Authenticity Results (Quantative Analysis)}
\autoref{fig:coverage-distribution} shows the distribution of feature coverage similarity of all the mined regressions.
We can see that the most regressions have a high feature coverage similarity (mean is 0.85),
indicating that the most regression features tested in the regression-inducing commit are likely to be preserved in the regression-fixing commit.

\paragraph{Authenticity Results (Qualitative Analysis)}
We confirm that 89 out of 90 regressions (each one third for high, medium, and low coverage similarity score) are authentic.
We observe that the authentic regressions with low coverage score are usually caused by
(1) code refactoring and
(2) deviated program execution leading to the coverage of new extra code.
As for the latter, the bug leads the program to execute other branches and, in turn, invoke new code,
causing a large $E_{ric}$ including many utility functions.
Nevertheless, in most cases, the core feature is still within $E_{rfc} \cap E_{ric}$.

\begin{figure}[t]
  \centering
  \includegraphics[scale=0.6]{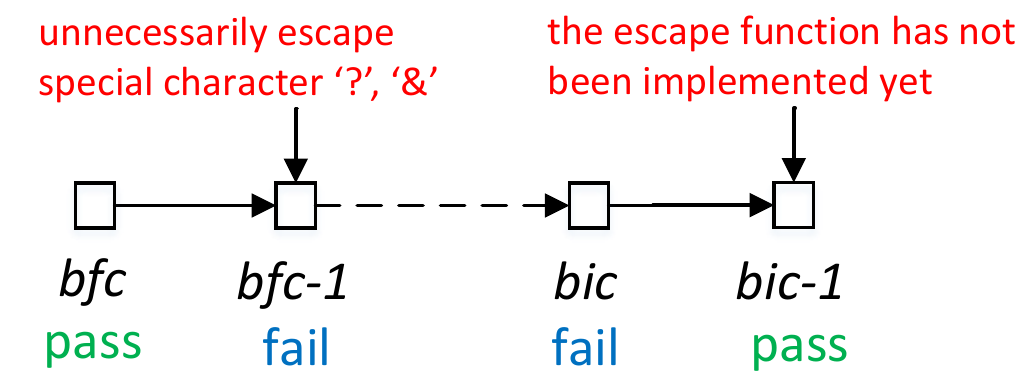}
  \caption{False positive reported by \tool}\label{fig:false-positive}
\end{figure}

The false positive is caused by specification change, with low feature coverage score of 0.043.
\autoref{fig:target-example} and \autoref{fig:false-positive} show the reported regression
where the feature is to parse a piece of HTML text into a Java object.
Given a piece of HTML text, if it contains a special character like `\&' or `?' in the value of \texttt{href} property,
the parser should not escape them to generate the Java object.
However, in \textit{bfc}-1, those characters are escaped,
which makes the test case fail.
\tool reports the revision \textit{bic}-1 which makes the test case pass.
However, our manual investigation shows that there is no escape functionality in \textit{bic}-1,
where only naive string matching is implemented.
Thus, the reported \textit{bic} accidentally passes the test case.
To fundamentally address the problem,
we need to fully extract the semantics of two executions between \textit{bfc} and \textit{bic}.
We leave it to our future work.

\begin{lstlisting}[caption={A regression test for checking whether special character in HTML can be handled well},
label={fig:target-example}, float=t]
@Test 
public void strictAttributeUnescapes() {
        String html = "<a href='?foo=bar&mid&lt=true'>O</a>";
        Elements els = Jsoup.parse(html).select("a");
        assertEquals("?foo=bar&mid&lt=true", els.first().attr("href"));
    }
}
\end{lstlisting}

\paragraph{Diversity Analysis}
\autoref{tab:diversity} shows that the regressions collected by \tool are more diverse than
Defects4j regarding the number of libraries, the involved library topics, and the types of exceptions.
Comparing to Defects4j, regressions in \tool use 7.3X libraries, cover 3.5X topics,
and raise 1.2X more types of exceptions.
More details of Venn Diagram to compare the diversity of \tool and Defects4j is available at \cite{regminer}.

\begin{table}[h]
\small
\centering
\tabcolsep=0.1cm
\caption{Diversity Evaluation, $H$ represents the entropy}
\label{tab:diversity}
\begin{tabular}{|l|l|ccc|cc|}
\hline
\multirow{2}{*}{} & \multirow{2}{*}{\textbf{\#Prj}} & \multicolumn{3}{c|}{\textbf{Library Diversity}}                                              & \multicolumn{2}{c|}{\textbf{Exception Diversity}}      \\ \cline{3-7}
                  &                                 & \multicolumn{1}{c|}{\textbf{\#Library}} & \multicolumn{1}{c|}{\textbf{\#Topic}} & \textbf{H} & \multicolumn{1}{c|}{\textbf{\#Exception}} & \textbf{H} \\ \hline
RegMiner          & \multicolumn{1}{c|}{147}        & \multicolumn{1}{c|}{818}                & \multicolumn{1}{c|}{110}              & 6.206      & \multicolumn{1}{c|}{56}                   & 1.221      \\ \hline
Defects4j         & \multicolumn{1}{c|}{17}         & \multicolumn{1}{c|}{122}                & \multicolumn{1}{c|}{31}               & 4.709      & \multicolumn{1}{c|}{68}                   & 0.883      \\ \hline
\end{tabular}
\end{table}

Overall, our conservative strategy (i.e., the strict conditions to report a regression regarding test compatibility on the passing and failing revisions) lead to high precision,
at the cost of moderate recall.
In our future work, we will study \tool's improvement on recall (see Section~\ref{sec:discussion} for more details).

\subsection{Ablation Study}
\subsubsection{Experiment Setup}
We disable, replace, and enable the three components respectively to evaluate their effectiveness.

\paragraph{Regression Potential}
Given that the number of regression fixing commits is $N$ ($N=$\regressionNum), we randomly select another $N$ bug fixing commits.
Then, we let \tool generate regression potential metric for each fixing commit,
comparing the metrics distribution of the two groups.
Moreover, we further evaluate our regression potential metrics by investigating the effectiveness of its rank on the regression-fixing commits.
Specifically, given $N$ regression-fixing commits and $\alpha N$ ($\alpha > 1$) non-regression fixing commits, we evaluate how many regression-fixing commits
\tool can report by looking the first $k$\% reported commits.
We let $\alpha > 1$ because that
non-regression fixing commits are usually more than regression-fixing commits.
We choose $\alpha=2$ in this study.

\paragraph{Test Dependency Migration and Validation Effort Minimization}
In this study, we disable and simplify the capability of test dependency migration and validation effort minimization separately.

\begin{itemize}[leftmargin=*]
  \item \textbf{Disabled Test Dependency Migration (\tool$_{\neg TDM}$):}
    We have \tool$_{\neg TDM}$ for \tool with test dependency migration disabled.
    In \tool$_{\neg TDM}$, we copy the identified test case into the target revision without dependency analysis.
  \item \textbf{Simplified Validation Effort Minimization (\tool\\$_{\neg VEM+bisec}$):}
    We have \tool$_{\neg VEM+bisec}$ for \tool with validation effort minimization simplified.
    In \tool$_{\neg VEM+bisec}$, we search for regressions by using git-bisect strategy \cite{git-bisect}.
    Moreover, if \tool$_{\neg VEM+bisec}$ visits a revision $c_{inv}$ without any feedback, we conservatively estimate a failure feedback on $c_{inv}$.
  \item \textbf{Disabled TDM and Simplified VEM (\tool$_{\neg TDM}$ + $bisec)$:}
    We have \tool$_{\neg TDM}$ + $bisec)$ for \tool with test dependency migration disabled and search for regression with git-bisect strategy.
  \item \textbf{Disabled TDM and Simplified VEM (\tool$_{\neg TDM}$ + \\$gitblame$):}
    We have \tool$_{\neg TDM}$ + $gitblame$ for \tool with test dependency migration disabled and search for regression with git-blame strategy on the difference between \textit{rfc} and \textit{rfc}-1.
    We consider the strategy on a regression successful if
    (1) git-blame can report \textit{ric} and
    (2) the test migration from \textit{rfc} to \textit{ric} is successful.
\end{itemize}


We compare \tool with the aforementioned variants on the datasets in both close-world and open-world experiments.
Given the open-world experiment takes \weekNum weeks
in this study, we replicate the experiment with the following simplification.
Given \tool reports $N$ regressions in the open-world experiment ($N=$\regressionNum in this study),
we randomly sample $2\times N$ bug-fixing commits, and
run \tool, \tool$_{\neg TDM}$, and \tool$_{\neg VEM}$ on the $3\times N$ bug fixing commits to
observe their performance.

\subsubsection{Results}\label{sec:rq3}

\begin{figure}[t]
  \centering
  \includegraphics[scale=0.22]{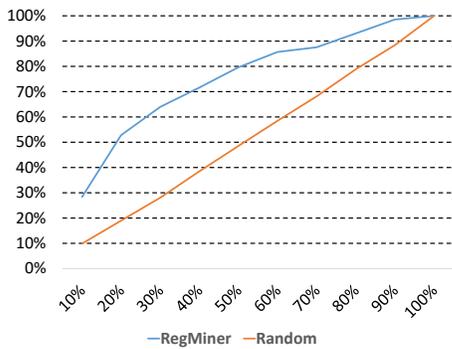}
  \caption{
  Effectiveness of regression potential metrics to select fixing commits with regression potential.
  We choose $\alpha=2$ i.e., the total number of fixing commits is $3\times N$ ($N$=1035) in which the number of regression-fixing commits is $N$.
  }
  \label{fig:regression-potential}
\end{figure}

\paragraph{Regression Potential Estimation}
We evaluate the regression potential scores on the \regressionNum regression-fixing commits and \regressionNum randomly chosen bug-fixing commits.
Overall, the regression-fixing group has an average of 0.51 and median of 0.43;
in contrast, the bug-fixing group has an average of 0.38 and median of 0.30.
We apply unpaired two-samples Wilcoxon test on two groups and have the $p$-value smaller than $10^{-5}$,
indicating that the regression-fixing group is significantly different from the bug-fixing group.
In addition, \autoref{fig:regression-potential} shows the effectiveness of \tool to select the regression-fixing commits.
Overall, the first 20\% ranked commits can include over 50\% regression-fixing commits (the blue curve in \autoref{fig:regression-potential}),
outperforming random selection (the orange curve in \autoref{fig:regression-potential}).
Thus, we conclude that our regression potential metric is effective.

\paragraph{Test Dependency Migration and Validation Effort Minimization}
\autoref{tab:performance} shows the overall regression retrieval performance.
Overall, comparing to the baselines, our solution shows its effectiveness.
While achieving the leading performance,
we also find that the variants such as \tool$_{\neg VEM+bisec}$ sometimes can report the regressions missed by \tool.
It is because different search strategy makes different tradeoffs,
leading to different search results.
In addition, simplified variants such as \tool$_{\neg TDM+bisec}$ and \tool$_{\neg TDM+gitblame}$ incur less runtime overhead, but have large degradation on recall.




\begin{table}[h]
\caption{
The overall performance of regression retrieval
}
\label{tab:performance}
\small
\tabcolsep=0.1cm
\centering
\begin{tabular}{|l|ccc|cc|}
\hline
\multirow{2}{*}{\textbf{Approach}} & \multicolumn{3}{c|}{\textbf{\begin{tabular}[c]{@{}c@{}}Close-world \\ Experiment\end{tabular}}} & \multicolumn{2}{c|}{\textbf{\begin{tabular}[c]{@{}c@{}}Open-world \\ Experiment\end{tabular}}} \\ \cline{2-6}
                                   & \multicolumn{1}{c|}{\textbf{Prec}}   & \multicolumn{1}{l|}{\textbf{Rec}}   & \textbf{Time (h)}  & \multicolumn{1}{l|}{\textbf{\#Reg}}                     & \textbf{Time (h)}                    \\ \hline
RegMiner                           & \multicolumn{1}{c|}{1.0}             & \multicolumn{1}{c|}{0.56}           & 4.29               & \multicolumn{1}{c|}{1035}                               & 135.38                               \\ \hline
RegMiner$_{\neg TDM}$              & \multicolumn{1}{c|}{1.0}             & \multicolumn{1}{c|}{0.32}           & 2.21               & \multicolumn{1}{c|}{604}                                & 64.29                                \\ \hline
RegMiner$_{\neg VEM+biset}$        & \multicolumn{1}{c|}{1.0}             & \multicolumn{1}{c|}{0.47}           & 2.74               & \multicolumn{1}{c|}{629}                                & 73.84                                \\ \hline
RegMiner$_{\neg TDM+biset}$        & \multicolumn{1}{c|}{1.0}             & \multicolumn{1}{c|}{0.04}           & 0.38               & \multicolumn{1}{c|}{114}                                & 16.65                                \\ \hline
RegMiner$_{\neg TDM+gitblame}$     & \multicolumn{1}{c|}{1.0}             & \multicolumn{1}{c|}{0.20}           & 0.85               & \multicolumn{1}{c|}{311}                                & 27.40                                \\ \hline
\end{tabular}
\end{table}

\subsection{Threats to Validity}
One internal threat is the precision (100\%) reported in our close-world experiment,
which is only reported on 50 regressions and 50 non-regression bugs.
Given the small number of regressions,
it may not be sufficient to conclude that the precision can be applied to the open-world experiment.
To mitigate the issue, we further manually sample 90 regressions in the open-world experiment,
to complement the precision evaluation.
One external threat lies in that our implementation is based on Java projects,
further studies on other popular programming language such as Python and C++ are still needed to generalize our findings.

\section{Discussion}\label{sec:discussion}
Our experiments have shown that \tool can successfully construct a runnable regression dataset,
with its size continuously growing, showing good potential as a research infrastructure to support
various software engineering studies.
In this section, we discuss the follow-up work regarding
(1) improving the mining effectiveness,
(2) realizing the potential of this infrastructure, and
(3) its alternative applications.

\noindent
\subsection{Mining Effective (Recall) Improvement}
We can improve the mining effectiveness by
(1) on-the-fly rewriting rule generation,
(2) regression test case selection, and
(3) multiple-test migration.

\noindent
\textbf{On-the-fly Rule Generation.}
\tool uses manually defined AST-rewriting rules to reduce the compilation errors induced by test-case migration.
In the future work, we will investigate how to derive semantic-preserving rules on-the-fly by comparing the different versions of library used in two commits.

\noindent
\textbf{Regression Test Selection.}
\tool only uses the test case created in the regression-fixing commit,
assuming that regressions are fixed along with a complementary regression test.
The conservative strategy may miss some regressions with their regression tests which have been created before the fixing-commit.
We will expand the scope of regression-test detection to improve the recall.

\noindent
\textbf{Multiple-test Migration.}
Migrating $N$ ($N > 1$) test cases potentially allows us to find multiple regressions by a single search.
The challenge lies in how to deal with different feedbacks of different tests when searching for the regression-inducing commit.
We will design a more optimal solution to maximize the “reward” of the search.

\subsection{Dataset Quality}
Since we intend to build a large dataset to evaluate various software testing, debugging, and repair research work,
the dataset quality is important.
Despite the chance of false positives is small, 
the open-world experiment still reveals such a possibility.
Moreover, despite existing techniques such as delta-debug\-ging \cite{zelleryesterday, wang2021probablisitic} can help \tool to isolate failure-inducing changes,
the human effort to verify and annotate the changes is almost inevitable with the existing software engineering techniques.

Therefore, 
we call for and foresee a crowdsourcing platform for researchers and volunteers to contribute,
with the shared \tool facility.
The following questions need to be answered:
\begin{enumerate}[leftmargin=*]
  \item \textbf{Social Aspect:} how to design a crowdsourcing system to involve the researchers in the community?
  \item \textbf{Tool Support:} how to design an intuitive bug/regression annotation tool to confirm the bug with minimum effort?
  \item \textbf{Technical Design:} how to improve existing software engineering techniques to recommend failure-inducing more precisely?
\end{enumerate}

We designed a regression annotation prototype as the first step towards this direction (see \cite{regminer}).
The demonstration are available at \url{https://regminer.github.io/}.






\subsection{Potential Beneficiary}
\tool is intended for the researchers in the PL/SE community,
which allows them to flexibly search for different types of bugs/regressions for their study.
Nevertheless, we foresee that \tool can also be used by practitioners in software industry to build project-specific benchmark.
Industrial practitioners can use \tool to
(1) understand what project functionalities are more likely to introduce regression bugs,
(2) who are more likely to introduce regression, and
(3) collect reusable regression patches for future fix recommendation and reference.
The first and second application can help project managers to make organizational decision, and
the third application is useful for software developers to avoid reinventing wheels. 
\section{Related Work}\label{sec:related_work}

\subsection{Bug Dataset Construction}
Researchers have proposed many bug datasets in the community.
Hutchins et. al \cite{hutchins1994experiments} construct the first real-world bug dataset.
Do, Elbaum, and Rothermel \cite{sir} further construct the SIR dataset.
Following their work,
various datasets are constructed from
programming assignments and competitions (e.g., Marmoset \cite{marmoset} QuixBugs \cite{quixbugs}, IntroClass \cite{introclass}, Codeflaws \cite{codeflaws}, etc.),
open source projects (e.g., DbgBench \cite{bohme2017bug}, Defects4j \cite{just2014defects4j}, BugsJS \cite{BugsJS}, Bugs.jar \cite{bugs.jar}, etc.), and
runtime continuous integration scenarios (e.g., BEARS \cite{bears} and BugSwarm \cite{bugswarm}).
The most relevant dataset CoREBench
\cite{bohme2014corebench}, which is a regression dataset of 70 C/C++ regressions.

Those bug datasets are constructed manually, which naturally affects the scalability and representativeness of the bug dataset.
Dallmeier and Zimmermann \cite{dallmeier2007extraction} make the first attempt to construct bug dataset in a semi-automatic way, via analyzing bug issues and their commits.
Zhao et. al \cite{zhao2019automatically} further propose to replicate bugs based on Android bug reports.
Recently,
BEARS \cite{bears} and BugSwarm \cite{bugswarm} are proposed to construct the bug dataset by collecting the buggy and the patched version on Continuous Integration system.
Following their work, Jiang et. al \cite{bugbuilder} further propose BugBuilder to construct dataset by isolating the bug-relevant changes.
\tool is different in that we target for a regression bug dataset,
and address new technical challenges such as regression-fixing commit prediction, test dependency migration, and validation effort minimization.

\subsection{Regression Research}
Regression research includes regression fault localization \cite{yi2015synergistic, wang2019explaining, hodovan2016modernizing, hodovan2017coarse, kiss2018hddr, misherghi2006hdd, zelleryesterday, zeller2002isolating, brummayer2009fuzzing, wang2021probablisitic},
regression testing \cite{labuschagne2017measuring, mezzetti2018type, godefroid2020differential, bertolino2020learning}, and
regression explanation \cite{yi2015synergistic, wang2019explaining}.
Zeller \cite{zelleryesterday} pioneers the delta debugging algorithm, which is followed by a number of variants in specific scenarios \cite{hodovan2016modernizing, hodovan2017coarse, kiss2018hddr, misherghi2006hdd}.
The most recent delta-debugging variant is proposed by Wang et. al \cite{wang2021probablisitic}, as probabilistic delta debugging, to further lower the algorithm complexity while improve the accuracy.
Tan and Roychoudhury \cite{tan2015relifix} propose a regression bug repair technique by searching over the regression revision through predefined code transformation rules.
As for regression explanation, Wang et. al \cite{wang2019explaining} propose a state-of-the-art regression explanation technique by proposing a novel alignment slicing algorithm on the execution traces of the regression version and past working version.
However, researchers usually either prepare their own dataset with limited size or inject mutated regressions which are less representative for the real-world regressions.

Our \tool solution can largely mitigate the challenges, laying foundation for the future regression analysis and
motivating both empirical and experimental facilities for the follow-up regression research.

\section{Conclusion}\label{sec:conclusion}
In this work, we propose \tool which can automatically construct regression dataset.
We address the challenges of
predicting regression-fixing commits,
migrating test and its dependencies, and
minimizing the regression validation effort.
Our close-world experiment shows that \tool achieves acceptable recall.
Our open-world experiment shows that \tool has constructed an authentic and diverse regression dataset within a short time.

\section*{Acknowledgement}
We sincerely thank anonymous reviewers for their comments to improve this work.
This research is supported in part by
the Minister of Education, Singapore (T2EP20120-0019,
T1-251RES1901, MOET32020-0004),
A*STAR, CISCO Systems (USA) Pte. Ltd and National University of Singapore under its Cisco-NUS Accelerated Digital Economy Corporate Laboratory (Award I21001E0002),
and National Natural Science Foundation of China (62172099). 

%
\bibliographystyle{ACM-Reference-Format}
\bibliography{sample-base}

%

\end{document}